# A multiple-relaxation-time lattice Boltzmann model for convection heat transfer in porous media

Qing Liu[1], Ya-Ling He[1,*], Qing Li[2], Wen-Quan Tao[1]

[1]Key Laboratory of Thermo-Fluid Science and Engineering of MOE, School of Energy and Power Engineering, Xi'an Jiaotong University, Xi'an, Shaanxi, 710049, P.R.China

[2]Energy Technology Research Group, Faculty of Engineering and the Environment, University of Southampton, Southampton SO17 1BJ, United Kingdom

**Abstract**

In this paper, a two-dimensional (2D) multiple-relaxation-time (MRT) lattice Boltzmann (LB) model is developed for simulating convection heat transfer in porous media at the representative elementary volume scale. In the model, a MRT-LB equation is used to simulate the flow field, while another MRT-LB equation is employed to simulate the temperature field. The effect of the porous media is considered by introducing the porosity into the equilibrium moments, and adding a forcing term to the MRT-LB equation of the flow field in the moment space. The present MRT-LB model is validated by numerical simulations of several 2D convection problems in porous media. The numerical results are in good agreement with the well-documented data reported in the literature.

## 1. Introduction

The analysis of convection heat transfer in porous media has attracted considerable attention due to its importance in the related technological and engineering applications, such as geothermal energy systems, chemical catalytic reactors, crude oil extraction, solar power collectors, electronic device cooling, and contaminant transport in groundwater. In the past several decades, various traditional numerical methods, such as the finite volume method, the finite difference method, and the finite element method, have been used to study convection heat transfer in finite porous enclosures. Comprehensive literature surveys of this subject can be found in Refs. [1, 2]. In particular, the non-Darcy effects on convection heat transfer in fluid-saturated porous media have been investigated numerically by many researchers [3-6].

The lattice Boltzmann (LB) method, which evolves from the lattice-gas automata (LGA) method [7], has gained great success in modeling complex fluid flows and simulating complex physics in fluids due to its kinetic background [8-11]. As a mesoscopic method based on the kinetic equation, the LB method has some distinctive merits over the traditional numerical methods (see, e.g., Ref. [12]). Owing to its kinetic nature and distinctive computational feature, the LB method has been successfully applied to study fluid flows in porous media soon after its emergence [13]. The existing LB models for porous flows can be generally classified into two categories, i.e., the pore scale method [13-15] and the representative elementary volume (REV) scale method [16-21]. In the pore scale method [13-15], the standard LB model is used to simulate the fluid flows in the pores, and the interaction between the solid and fluid phases is realized by using the no-slip bounce-back rule. The detailed flow information of the pores can be obtained by this method, which can be utilized to investigate macroscopic relations. However, this method needs the detailed geometric information of the pores, and each pore needs several lattice nodes in simulations, so the computation domain is small because of the limited computer resources [19]. In the REV scale LB method [16-18], an additional term is added to the standard LB equation to consider the effect of the porous media based on some semiempirical models, such as the Darcy model, the Brinkman-extended Darcy model and the Forchheimer-extended Darcy model. Based on the so-called Brinkman-Forchheimer-extended Darcy model (also called the generalized model) [5, 34] that overcomes some limitations of the Darcy model, Brinkman-extended Darcy model and Forchheimer-extended Darcy model, Guo and Zhao [19] proposed a generalized LB model for simulating isothermal incompressible porous flows. In this model, the porosity is included into the equilibrium distribution function, and a forcing term is added to the LB equation to account for the Darcy (linear) and Forchheimer (nonlinear) drag forces of the solid matrix. Subsequently, Guo and Zhao [20] extended the generalized LB model to incompressible thermal flows in fluid-saturated porous media by using the double-distribution-function (DDF) approach. In the literature, Seta et al. [21] confirmed the reliability and the computational efficiency of the LB method in studying natural convection flow in porous media with the generalized model. As reported in Ref. [21], the LB method needs less computational time than the finite difference method to obtain the same accurate solutions of natural convection flow in porous media on the same grid size.

However, to the best of our knowledge, the existing LB models for porous flows at the REV scale employ the Bhatnagar-Gross-Krook (BGK) collision model [22] to represent the collision process in the evolution equation. Although the lattice Bhatnagar-Gross-Krook (LBGK) model has become the most popular one for its extreme simplicity, it has also received several well-known criticisms, among which the most important one is the numerical instability at low viscosities. Fortunately, it has been demonstrated that the shortcomings of the LBGK model can be resolved by using the multiple-relaxation-time (MRT) collision model [23, 24]. In the LB community, it has been widely accepted that the MRT model can improve the numerical stability by separating the relaxation rates of the hydrodynamic (conserved) and nonhydrodynamic (nonconserved) moments [25-27]. In recent years, the MRT-LB method has been successfully applied to simulate complex fluid flows such as axisymmetric flows [28], microscale flows [29], and multiphase flows [30, 31]. In Refs. [32, 33], the

MRT-LB method has been used to study two-dimensional convective flows in the absence of porous media. The basic idea of the MRT-LB method [32, 33] is that the flow and temperature fields are solved separately by two different MRT-LB equations. As reported in Ref. [33], the MRT-LB model with two MRT-LB equations is second-order accurate for simulating incompressible thermal flows with the Boussinesq approximation.

In this paper, we aim to develop a MRT-LB model for simulating convection heat transfer in fluid-saturated porous media at the REV scale, which can be viewed as an extension to some previous studies. The rest of this paper is organized as follows. The macroscopic governing equations are described in Section 2. The MRT-LB model is proposed in Section 3. In Section 4, the numerical tests are presented. Finally, a brief conclusion is made in Section 5.

## 2. Macroscopic governing equations

The flow is supposed to be two-dimensional, laminar and incompressible with negligible viscous heat dissipation. In addition, it is assumed that the solid phase is in local thermal equilibrium with the fluid phase [2]. Based on the generalized model, the dimensional governing equations for convection heat transfer in a homogeneous, isotropic and fluid-saturated porous medium at the REV scale can be written as [5, 6, 34]:

$$\nabla \cdot \mathbf{u} = 0, \tag{1}$$

$$\frac{\partial \mathbf{u}}{\partial t} + (\mathbf{u} \cdot \nabla)\left(\frac{\mathbf{u}}{\phi}\right) = -\frac{1}{\rho_0}\nabla(\phi p) + \upsilon_e \nabla^2 \mathbf{u} + \mathbf{F}, \tag{2}$$

$$\sigma \frac{\partial T}{\partial t} + \mathbf{u} \cdot \nabla T = \nabla \cdot (\alpha_e \nabla T) + Q''', \tag{3}$$

where $\rho_0$ is the mean fluid density, $\mathbf{u}$, $T$, and $p$ are the volume-averaged fluid velocity, temperature, and pressure, respectively; $\phi$ is the porosity, $K$ is the permeability, $\upsilon_e$ is the effective kinetic viscosity, and $Q'''$ is the internal heat source term; the coefficient $\sigma = \left[\phi(\rho_f c_{pf}) + (1-\phi)(\rho_s c_{ps})\right]/(\rho_f c_{pf})$ is the thermal capacity ratio, in which $\rho_f$ ($\rho_s$) and $c_{pf}$ ($c_{ps}$) are the density and specific heat of the fluid (solid) phase, respectively; the effective thermal diffusivity $\alpha_e = k_e/(\rho_f c_{pf})$, in which $k_e$ is the effective thermal conductivity. **F** denotes the total body force induced by the porous media and other external force, which can be expressed as [34]

$$\mathbf{F} = -\frac{\phi \upsilon}{K}\mathbf{u} - \frac{\phi F_\phi}{\sqrt{K}}|\mathbf{u}|\mathbf{u} + \phi \mathbf{G}, \tag{4}$$

where $\upsilon$ is the kinetic viscosity, $|\mathbf{u}| = \sqrt{u_x^2 + u_y^2}$, in which $u_x$ and $u_y$ are the components of **u** in the *x*- and *y*-directions, respectively. According to the Boussinesq approximation, the buoyancy force **G** induced by the gravity is given by $\mathbf{G} = g\beta(T-T_0)\mathbf{j}$, where $g$ is the gravitational acceleration, $\beta$ the thermal expansion coefficient, $T_0$ the reference temperature, and **j** the unit vector in the *y*-direction.

Based on Ergun's relation [35], the geometric function $F_\phi$ and the permeability $K$ can be expressed as [36]

$$F_\phi = \frac{1.75}{\sqrt{150\phi^3}}, \quad K = \frac{\phi^3 d_p^2}{150(1-\phi)^2}, \tag{5}$$

where $d_p$ represents the solid particle diameter.

Several dimensionless parameters characterize the system governed by Eqs. (1)-(3): the thermal Rayleigh number $Ra$, the internal Rayleigh number $Ra_I$ (for thermal convection flows with internal heat source), the Darcy number $Da$, the viscosity ratio $J$, the Prandtl number $Pr$, which are defined as follows

$$Ra = \frac{g\beta\Delta T L^3}{\upsilon \alpha_e}, \quad Ra_I = \frac{g\beta Q''' L^5}{\upsilon \alpha_e^2}, \quad Da = \frac{K}{L^2}, \quad Pr = \frac{\upsilon}{\alpha_e}, \quad J = \frac{\upsilon_e}{\upsilon},$$

where $L$ and $\Delta T$ are the characteristic length and temperature, respectively.

## 3. MRT-LB model for thermal flows in porous media

### 3.1 MRT-LB equation for the flow field

For the flow field, a LB equation with the MRT collision model [23-27] is considered in this study. According to Refs. [28, 30], the MRT-LB equation with an explicit treatment of the forcing term can be written as

$$\mathbf{f}(\mathbf{x}_k + \mathbf{e}\delta_t, t_n + \delta_t) - \mathbf{f}(\mathbf{x}_k, t_n) = -\mathbf{M}^{-1}\mathbf{\Lambda}\left[\mathbf{m} - \mathbf{m}^{(eq)}\right]\bigg|_{(\mathbf{x}_k, t_n)} + \mathbf{M}^{-1}\delta_t\left(\mathbf{I} - \frac{\mathbf{\Lambda}}{2}\right)\mathbf{S}, \quad (6)$$

where $\mathbf{M}$ is a $Q \times Q$ orthogonal transformation matrix ($Q$ represents the number of discrete velocities), $\mathbf{\Lambda} = \mathbf{M}\tilde{\mathbf{\Lambda}}\mathbf{M}^{-1} = \text{diag}(s_0, s_1, \cdots, s_{Q-1})$ is a diagonal relaxation matrix, in which $\tilde{\mathbf{\Lambda}}$ is the collision matrix and $\{s_i \mid 0 \leq i \leq Q-1\}$ are relaxation rates, and $\mathbf{I}$ is the identity matrix. The boldface symbols, $\mathbf{f}$, $\mathbf{m}$, $\mathbf{m}^{(eq)}$, and $\mathbf{S}$ denote $Q$-dimensional column vectors:

$$\mathbf{f}(\mathbf{x}_k, t_n) = \left(f_0(\mathbf{x}_k, t_n), f_1(\mathbf{x}_k, t_n), \cdots, f_{Q-1}(\mathbf{x}_k, t_n)\right)^{\text{T}},$$

$$\mathbf{f}(\mathbf{x}_k + \mathbf{e}\delta_t, t_n + \delta_t) = \left(f_0(\mathbf{x}_k + \mathbf{e}_0\delta_t, t_n + \delta_t), \cdots, f_{Q-1}(\mathbf{x}_k + \mathbf{e}_{Q-1}\delta_t, t_n + \delta_t)\right)^{\text{T}}$$

$$\mathbf{m}(\mathbf{x}_k, t_n) = \left(m_0(\mathbf{x}_k, t_n), m_1(\mathbf{x}_k, t_n), \cdots, m_{Q-1}(\mathbf{x}_k, t_n)\right)^{\text{T}}$$

$$\mathbf{m}^{(eq)}(\mathbf{x}_k, t_n) = \left(m_0^{(eq)}(\mathbf{x}_k, t_n), m_1^{(eq)}(\mathbf{x}_k, t_n), \cdots, m_{Q-1}^{(eq)}(\mathbf{x}_k, t_n)\right)^{\text{T}},$$

$$\mathbf{S} = \left(S_0, S_1, S_2, \cdots, S_{Q-1}\right)^{\text{T}}, \quad (7)$$

where T is the transpose operator, $f_i(\mathbf{x}_k, t_n)$ is the volume-averaged discrete distribution function with velocity $\mathbf{e}_i$ at lattice node $\mathbf{x}_k$ and discrete time $t_n$ (at REV scale), $\mathbf{m}(\mathbf{x}_k, t_n)$ and $\mathbf{m}^{(eq)}(\mathbf{x}_k, t_n)$ are the velocity moments and the corresponding equilibrium moments, respectively, and $S_i$ is the component of the forcing term $\mathbf{S}$.

For fluid flows through porous media in two dimensions, we use the D2Q9 lattice model, of which the nine discrete velocities $\{\mathbf{e}_i \mid i = 0, 1, \cdots, 8\}$ are given by [39]

$$\mathbf{e}_i = \begin{cases} (0,0), & i=0 \\ \left(\cos\left[(i-1)\pi/2\right], \sin\left[(i-1)\pi/2\right]\right)c, & i=1 \sim 4 \\ \left(\cos\left[(2i-9)\pi/4\right], \sin\left[(2i-9)\pi/4\right]\right)\sqrt{2}c, & i=5 \sim 8 \end{cases} \tag{8}$$

where $c = \delta_x/\delta_t$ is the lattice speed, in which $\delta_t$ and $\delta_x$ are the discrete time step and lattice spacing, respectively. In the present MRT-LB model, $\delta_x$ is set equal to $\delta_t$, which leads to $c=1$ (in lattice unit).

The transformation matrix $\mathbf{M}$ linearly maps the discrete distribution functions $\mathbf{f} \in \mathbb{V} = \mathbb{R}^9$ (velocity space) to their velocity moments $\mathbf{m} \in \mathbb{M} = \mathbb{R}^9$ (moment space):

$$\mathbf{m} = \mathbf{Mf}, \quad \mathbf{f} = \mathbf{M}^{-1}\mathbf{m}. \tag{9}$$

The nine velocity moments are:

$$\mathbf{m} = (m_0, m_1, m_2, m_3, m_4, m_5, m_6, m_7, m_8)^T$$
$$= (\rho, e, \varepsilon, j_x, q_x, j_y, q_y, p_{xx}, p_{xy})^T, \tag{10}$$

where $m_0 = \rho$ is the fluid density, $m_1 = e$ is related to energy, $m_2 = \varepsilon$ is related to energy square, $m_{3,5} = j_{x,y}$ are components of the momentum $\mathbf{J} = (j_x, j_y) = \rho\mathbf{u}$, $m_{4,6} = q_{x,y}$ are related to energy flux, and $m_{7,8} = p_{xx,xy}$ correspond to the symmetric and traceless components of the strain-rate tensor [25]. With the ordering of the above specified velocity moments, the transformation matrix $\mathbf{M}$ is given by ($c=1$) [25]

$$\mathbf{M} = \begin{pmatrix} 1 & 1 & 1 & 1 & 1 & 1 & 1 & 1 & 1 \\ -4 & -1 & -1 & -1 & -1 & 2 & 2 & 2 & 2 \\ 4 & -2 & -2 & -2 & -2 & 1 & 1 & 1 & 1 \\ 0 & 1 & 0 & -1 & 0 & 1 & -1 & -1 & 1 \\ 0 & -2 & 0 & 2 & 0 & 1 & -1 & -1 & 1 \\ 0 & 0 & 1 & 0 & -1 & 1 & 1 & -1 & -1 \\ 0 & 0 & -2 & 0 & 2 & 1 & 1 & -1 & -1 \\ 0 & 1 & -1 & 1 & -1 & 0 & 0 & 0 & 0 \\ 0 & 0 & 0 & 0 & 0 & 1 & -1 & 1 & -1 \end{pmatrix}. \tag{11}$$

Among the nine velocity moments $\{m_i \mid i = 0, 1, \cdots, 8\}$, only the density $m_0 = \rho$ and momentum

$m_{3,5} = j_{x,y}$ are conserved (hydrodynamic) quantities, while the other six velocity moments are non-conserved (kinetic) quantities. To include the influence of the porous media, the equilibrium moments $\mathbf{m}^{(eq)}$ for the non-conserved moments $\mathbf{m}$ can be defined as:

$$e^{(eq)} = -2\rho + \frac{3\rho|\mathbf{u}|^2}{\phi}, \quad \varepsilon^{(eq)} = \rho - \frac{3\rho|\mathbf{u}|^2}{\phi}$$

$$q_x^{(eq)} = -\rho u_x, \quad q_y^{(eq)} = -\rho u_y,$$

$$p_{xx}^{(eq)} = \frac{\rho(u_x^2 - u_y^2)}{\phi}, \quad p_{xy}^{(eq)} = \frac{\rho u_x u_y}{\phi}. \tag{12}$$

For incompressible fluid flows through porous media considered in this study, an incompressible MRT-LB equation is proposed to solve the flow field. To do so, the equilibrium distribution function of the LBGK model [38] for incompressible Navier-Stokes equation is employed. To account for the porosity $\phi$ of the porous media, the equilibrium distribution function is now modified as:

$$f_i^{(eq)} = \begin{cases} \rho_0 - (1-\omega_0)\frac{\phi p}{c_s^2} + \rho_0 s_0(\mathbf{u}), & i = 0 \\ \omega_i \frac{\phi p}{c_s^2} + \rho_0 s_i(\mathbf{u}), & i = 1 \sim 8 \end{cases}, \tag{13}$$

where $\{\omega_i \mid i = 0, 1, \cdots, 8\}$ are weight coefficients with $\omega_0 = 4/9$, $\omega_i = 1/9$ for $i = 1 \sim 4$, $\omega_i = 1/36$ for $i = 5 \sim 8$, $c_s = c/\sqrt{3}$ is the sound speed, and

$$s_i(\mathbf{u}) = \omega_i \left[ \frac{\mathbf{e}_i \cdot \mathbf{u}}{c_s^2} + \frac{(\mathbf{e}_i \cdot \mathbf{u})^2}{2\phi c_s^4} - \frac{|\mathbf{u}|^2}{2\phi c_s^2} \right], \tag{14}$$

Through the transformation matrix $\mathbf{M}$, the equilibrium distribution function $f_i^{(eq)}$ in the velocity space can be projected onto the moment space with $\mathbf{m}^{(eq)} = \mathbf{M}\mathbf{f}^{(eq)}$, where $\mathbf{f}^{(eq)} = \left(f_0^{(eq)}, f_1^{(eq)}, \cdots, f_8^{(eq)}\right)^T$. Based on the equilibrium distribution function $f_i^{(eq)}$ given by Eq. (13), the equilibrium moments $\mathbf{m}^{(eq)}$ for the velocity moments $\mathbf{m}$ can be obtained as

$$\mathbf{m}^{(eq)} = \left(\rho_0, e^{(eq)}, \varepsilon^{(eq)}, j_x, q_x^{(eq)}, j_y, q_y^{(eq)}, p_{xx}^{(eq)}, p_{xy}^{(eq)}\right)^T. \tag{15}$$

where

$$e^{(eq)} = -4\rho_0 + 6\phi p + \frac{3\rho_0 |\mathbf{u}|^2}{\phi}, \quad \varepsilon^{(eq)} = 4\rho_0 - 9\phi p - \frac{3\rho_0 |\mathbf{u}|^2}{\phi}$$

$$q_x^{(eq)} = -\rho_0 u_x, \quad q_y^{(eq)} = -\rho_0 u_y,$$

$$p_{xx}^{(eq)} = \frac{\rho_0 (u_x^2 - u_y^2)}{\phi}, \quad p_{xy}^{(eq)} = \frac{\rho_0 u_x u_y}{\phi}. \tag{16}$$

The incompressibility approximation, i.e., $\rho \approx \rho_0$ and $\mathbf{J} = (j_x, j_y) \approx \rho_0 \mathbf{u}$, have been used in the equilibrium moments of Eq. (15). In simulations, the mean fluid density $\rho_0$ is usually set to be 1 for simplicity.

The evolution of the MRT-LB equation (6) consists of two steps, i.e., the collision step and streaming step [25]. The collision step is executed in the moment space:

$$\mathbf{m}^+ = \mathbf{m} - \mathbf{\Lambda}\left[\mathbf{m} - \mathbf{m}^{(eq)}\right] + \delta_t \left(\mathbf{I} - \frac{\mathbf{\Lambda}}{2}\right)\mathbf{S}, \tag{17}$$

where $\mathbf{I}$ is the $9 \times 9$ identity matrix, and $\mathbf{S} = \mathbf{M}\tilde{\mathbf{S}}$, in which $\tilde{\mathbf{S}} = (\tilde{S}_0, \tilde{S}_1, \cdots, \tilde{S}_8)^\mathrm{T}$. The streaming step is still carried out in the velocity space:

$$f_i(\mathbf{x}_k + \mathbf{e}_i \delta_t, t_n + \delta_t) = f_i^+(\mathbf{x}_k, t_n), \tag{18}$$

where $\mathbf{f}^+ = \mathbf{M}^{-1}\mathbf{m}^+$. The components of the forcing term $\mathbf{S}$ in the moment space are given explicitly by

$$S_0 = 0, \quad S_1 = \frac{6\rho_0 \mathbf{u} \cdot \mathbf{F}}{\phi}, \quad S_2 = -\frac{6\rho_0 \mathbf{u} \cdot \mathbf{F}}{\phi}, \quad S_3 = \rho_0 F_x, \quad S_4 = -\rho_0 F_x,$$

$$S_5 = \rho_0 F_y, \quad S_6 = -\rho_0 F_y, \quad S_7 = \frac{2\rho_0 (u_x F_x - u_y F_y)}{\phi}, \quad S_8 = \frac{\rho_0 (u_x F_y + u_y F_x)}{\phi}, \tag{19}$$

where $\mathbf{F} = (F_x, F_y)$ is given by Eq. (4).

The diagonal relaxation matrix $\mathbf{\Lambda}$ is given by:

$$\mathbf{\Lambda} = \mathrm{diag}(s_0, s_1, s_2, s_3, s_4, s_5, s_6, s_7, s_8)$$

$$= \mathrm{diag}(1, s_e, s_\varepsilon, 1, s_q, 1, s_q, s_\upsilon, s_\upsilon). \tag{20}$$

The macroscopic fluid velocity $\mathbf{u}$ is defined as [19]

$$\mathbf{u} = \frac{\mathbf{v}}{l_0 + \sqrt{l_0^2 + l_1 |\mathbf{v}|}}, \tag{21}$$

where $\mathbf{v}$ is a temporal velocity

$$\mathbf{v} = \sum_{i=0}^{8} \mathbf{e}_i f_i / \rho_0 + \frac{\delta_t}{2} \phi \mathbf{G}. \tag{22}$$

The two parameters $l_0$ and $l_1$ in Eq. (21) are given by

$$l_0 = \frac{1}{2}\left(1 + \phi \frac{\delta_t}{2} \frac{\upsilon}{K}\right), \quad l_1 = \phi \frac{\delta_t}{2} \frac{F_\phi}{\sqrt{K}}.$$

The macroscopic fluid pressure $p$ is defined as

$$p = \frac{c_s^2}{\phi(1-\omega_0)}\left[\sum_{i=1}^{8} f_i + \rho_0 s_0(\mathbf{u})\right]. \tag{23}$$

The effective kinetic viscosity $\upsilon_e$ is defined as

$$\upsilon_e = c_s^2 \left(\tau_\upsilon - \frac{1}{2}\right)\delta_t \tag{24}$$

with $s_7 = s_8 = s_\upsilon = 1/\tau_\upsilon$. Note that when the nine relaxation rates are all equal, i.e., $\Lambda = (1/\tau_\upsilon)\mathbf{I}$, the MRT-LB equation (6) reduces to the LBGK equation with the equilibrium distribution function given by Eq. (13).

Through the Chapman-Enskog analysis [30, 39] of the evolution equation (6) in the moment space, the generalized Navier-Stokes equations (1) and (2) can be obtained exactly without artificial compressible error. Note that as $\phi \to 1$, the MRT-LB equation (6) (without the forcing term) reduces to the MRT-LB equation [40] for incompressible flows without porous media. Moreover, if we set $F_\phi = 0$ in the present MRT-LB equation, a simplified MRT-LB equation can be derived for the Brinkman-extended Darcy model.

**3.2 MRT-LB equation for the temperature field**

For the temperature field, the MRT-LB equation with a source term based on the D2Q5 lattice is defined as

$$\mathbf{g}(\mathbf{x}_k + \mathbf{e}\delta_t, t_n + \delta_t) - \mathbf{g}(\mathbf{x}_k, t_n) = -\mathbf{N}^{-1}\Theta\left[\mathbf{n} - \mathbf{n}^{(eq)}\right]\Big|_{(\mathbf{x}_k, t_n)} + \mathbf{N}^{-1}\delta_t \Psi, \tag{25}$$

where $\mathbf{N}$ is a $5\times 5$ orthogonal transformation matrix, and $\Theta$ is a diagonal relaxation matrix. The boldface symbols, $\mathbf{g}$, $\mathbf{n}$, $\mathbf{n}^{(eq)}$, and $\Psi$ are 5-dimensional column vectors, e.g., $\mathbf{g}(\mathbf{x}_k, t_n) = (g_0(\mathbf{x}_k, t_n), g_1(\mathbf{x}_k, t_n), \cdots, g_4(\mathbf{x}_k, t_n))^{\mathrm{T}}$, where $g_i(\mathbf{x}_k, t_n)$ is the temperature distribution function at lattice node $\mathbf{x}_k$ and discrete time $t_n$, and $\{\Psi_i \,|\, i = 0, 1, \cdots, 4\}$ are components of the source term $\Psi$.

In the D2Q5 model, the five discrete velocities $\{\mathbf{e}_i \,|\, i = 0, 1, \cdots, 4\}$ are given by

$$\mathbf{e}_i = \begin{cases} (0,0), & i = 0 \\ \left(\cos\left[(i-1)\pi/2\right], \sin\left[(i-1)\pi/2\right]\right)c, & i = 1 \sim 4 \end{cases}, \tag{26}$$

The transformation matrix $\mathbf{N}$ linearly transforms the discrete distribution functions $\mathbf{g} \in \mathbb{V} = \mathbb{R}^5$ to their velocity moments $\mathbf{n} \in \mathbb{M} = \mathbb{R}^5$:

$$\mathbf{n} = \mathbf{N}\mathbf{g}, \quad \mathbf{g} = \mathbf{N}^{-1}\mathbf{n}. \tag{27}$$

For the D2Q5 model, the transformation matrix $\mathbf{N}$ is given by [32, 33]:

$$\mathbf{N} = \begin{pmatrix} 1 & 1 & 1 & 1 & 1 \\ 0 & 1 & 0 & -1 & 0 \\ 0 & 0 & 1 & 0 & -1 \\ -4 & 1 & 1 & 1 & 1 \\ 0 & 1 & -1 & 1 & -1 \end{pmatrix}. \tag{28}$$

In the system of $\{g_i\}$, only the temperature $T \equiv n_0 = \sum_i g_i$ is conserved quantity. The equilibrium moments $\{n_i^{(eq)} \,|\, i = 0, 1, \cdots, 4\}$ for the velocity moments $\{n \,|\, i = 0, 1, \cdots, 4\}$ are defined as

$$n_0^{(eq)} = T, \quad n_1^{(eq)} = \frac{u_x T}{\sigma}, \quad n_2^{(eq)} = \frac{u_y T}{\sigma}, \quad n_3^{(eq)} = \varpi T, \quad n_4^{(eq)} = 0, \tag{29}$$

where $\varpi$ is a constant. As reported in Ref. [33], to avoid the so-called "checkerboard" type instability, $\varpi$ must be smaller than 1.

The components of the source term $\Psi$ are given as follows:

$$\Psi_0 = \frac{Q'''}{\sigma}, \quad \Psi_1 = 0, \quad \Psi_2 = 0, \quad \Psi_3 = \frac{\varpi Q'''}{\sigma}, \quad \Psi_4 = 0. \tag{30}$$

The diagonal relaxation matrix $\Theta$ is given by:

$$\Theta = \text{diag}(1, \zeta_1, \zeta_2, \zeta_3, \zeta_4). \tag{31}$$

Through the Chapman-Enskog analysis (see Appendix for details), we obtain the following macroscopic equation:

$$\sigma \frac{\partial T}{\partial t} + \mathbf{u} \cdot \nabla T = \nabla \cdot (\alpha_e \nabla T) + Q''' - \epsilon \frac{\delta_t}{2} \frac{\partial Q'''}{\partial t_1}. \tag{32}$$

Note that the last term $-\epsilon(\delta_t/2)\partial_{t_1} Q'''$ in the right-hand side of Eq. (32) can be neglected if the heat source term is time independent. In the LBGK model [41] for convection-diffusion equation with a heat source term, such an additional term has been removed by adding a source term to the evolution equation, and using the redefinition of the temperature. In order to remove the additional term in Eq. (32), in the present study we employ the following D2Q5 MRT-LB equation to solve the temperature field:

$$\mathbf{g}(\mathbf{x}_k + \mathbf{e}\delta_t, t_n + \delta_t) - \mathbf{g}(\mathbf{x}_k, t_n) = -\mathbf{N}^{-1}\Theta\left[\mathbf{n} - \mathbf{n}^{(\text{eq})}\right]\Big|_{(\mathbf{x}_k, t_n)} + \mathbf{N}^{-1}\delta_t\left(\tilde{\mathbf{I}} - \frac{\Theta}{2}\right)\Psi, \tag{33}$$

which also consists of two steps, i.e., the collision step and streaming step. The collision step is implemented in the moment space

$$\mathbf{n}^+ = \mathbf{n} - \Theta\left[\mathbf{n} - \mathbf{n}^{(\text{eq})}\right] + \delta_t\left(\tilde{\mathbf{I}} - \frac{\Theta}{2}\right)\Psi, \tag{34}$$

where $\tilde{\mathbf{I}}$ is a $5 \times 5$ identity matrix. The streaming step is carried out in the velocity space

$$g_i(\mathbf{x}_k + \mathbf{e}_i\delta_t, t_n + \delta_t) = g_i^+(\mathbf{x}_k, t_n), \tag{35}$$

where $\mathbf{g}^+ = \mathbf{N}^{-1}\mathbf{n}^+$.

The macroscopic temperature $T$ is now defined as

$$T = \sum_{i=0}^{4} g_i + \frac{\delta_t}{2}\frac{Q'''}{\sigma}. \tag{36}$$

Through the Chapman-Enskog analysis of the D2Q5 MRT-LB equation (33) in the moment space,

the temperature equation (3) can be recovered. The heat source term $Q'''$ considered in this study is a linear function of $T$, so the macroscopic temperature $T$ can be obtained from Eq. (36).

If the five relaxation rates are all equal, i.e., $\Theta = (1/\tau_T)\tilde{\mathbf{I}}$, the MRT-LB equation (33) reduces to the LBGK equation with the following equilibrium distribution function:

$$g_i^{(eq)} = \tilde{\omega}_i T \left(1 + \frac{\mathbf{e}_i \cdot \mathbf{u}}{\sigma c_{sT}^2}\right), \tag{37}$$

where the weight coefficients $\{\tilde{\omega}_i \mid i = 0, 1, \cdots, 4\}$ are given by $\tilde{\omega}_0 = (1-\varpi)/5$, $\tilde{\omega}_i = (4+\varpi)/20$ for $i = 1 \sim 4$, and $c_{sT}^2 = (4+\varpi)c^2/10$ is the sound speed of the D2Q5 model ($c_{sT}^2 = \sum_i \tilde{\omega}_i e_{ix}^2 = \sum_i \tilde{\omega}_i e_{iy}^2$, see Ref. [42]). The effective thermal diffusivity $\alpha_e$ can be written as $\alpha_e = \sigma c_{sT}^2 (\tau_T - 0.5)\delta_t$.

## 4. Numerical tests

In this section, numerical simulations of several 2D convection problems in porous media are carried out to demonstrate the effectiveness of the present MRT-LB model. For different boundary conditions in the systems of $\{f_i\}$ and $\{g_i\}$, the non-equilibrium extrapolation scheme [43] is adopted in this study. If needed, some other scheme [33] can also be adopted in the present MRT-LB model. Unless otherwise stated, we set $\rho_0 = 1$, $\varpi = -2$, $J = 1$, $\sigma = 1$, and $\delta_t = \delta_x = \delta_y = 1$ in all computations. The relaxation rates, $\{s_i \mid 0 \leq i \leq 8\}$ and $\{\zeta_i \mid 0 \leq i \leq 4\}$, are chosen as follows: $s_0 = s_3 = s_5 = 1$, $s_1 = s_2 = 1.1$, $s_4 = s_6 = 1.2$, $s_7 = s_8 = 1/\tau_\upsilon$, $\zeta_0 = 1$, $\zeta_1 = \zeta_2 = 1/\tau_T$, and $\zeta_3 = \zeta_4 = 1.5$. For convection problems in porous media, the dimensionless relaxation times $\tau_\upsilon$ and $\tau_T$ can be fully determined in terms of $Pr$, $Ra$ and $Ma$ [11, 44, 45], where $Ma = U/c_s$ is the Mach number ($U = \sqrt{\beta g \Delta T L}$ is the characteristic velocity). According to Ref. [11, 45], the dimensionless relaxation times for the flow field and temperature field can be determined as

$$\tau_\upsilon = \frac{1}{2} + \frac{MaJL}{c_s \delta_t}\sqrt{\frac{Pr}{Ra}}, \quad \tau_T = \frac{1}{2} + \frac{c_s^2(\tau_\upsilon - 0.5)}{\sigma c_{sT}^2 Pr}, \tag{37}$$

respectively. To comply with the stability criterion on $\upsilon_e$ and also the incompressible limit of the flow, the Mach number should be small (usually $Ma < 0.3$). In this study, $Ma$ is set to be 0.1 in all simulations.

4.1 Natural convection in a porous cavity

Natural convection flow in a fluid-saturated porous cavity has been investigated extensively by many researchers using traditional numerical methods [3-5] and LB method [20, 21]. The geometry and boundary conditions are illustrated in Fig. 1. The horizontal walls are thermally isolated, while the left and right vertical walls are kept at constant but different temperatures $T_h$ and $T_c$, respectively ($T_h > T_c$). $H$ and $L$ are the height and width of the cavity (aspect ratio $A = H/L = 1$), respectively. $\Delta T = T_h - T_c$ is the temperature difference (characteristic temperature), and $T_0 = (T_h + T_c)/2$ is the reference temperature. According to Ref. [45], the average Nusselt number $\overline{Nu}$ of the left (or right) vertical wall is defined as

$$\overline{Nu} = \int_0^L Nu(y)\,dy/L, \tag{38}$$

where $Nu(y) = -L(\partial T/\partial x)_{wall}/\Delta T$ is the local Nusselt number. The five-point formula is employed to calculate the temperature gradient $\partial T/\partial x$.

In this subsection, we first consider the case in which $\phi \to 1$ and $Da$ tends to infinity with $10^3 \leq Ra \leq 10^7$. Simulations for $Pr = 0.71$, $\phi = 0.9999$, and $Da = 10^8$ with $10^3 \leq Ra \leq 10^7$ using the present MRT-LB model are performed. The grid sizes of $150 \times 150$ for $Ra = 10^3$, $10^4$, and $250 \times 250$ for $Ra = 10^5$, $10^6$, $10^7$, are used in our simulations. The predicted average Nusselt numbers of the right vertical wall by the present MRT-LB model are listed in Table 1 and compared to the results from previous numerical studies [44-46]. As shown in Table 1, our results agree well with those of previous numerical studies.

We now present the results for various values of $Ra$, $Da$, and $\phi$. In simulations, $Pr$ is set to be 1, the grid sizes of $120\times120$, $200\times200$, and $250\times250$ are employed for $Da=10^{-2}$, $10^{-4}$, and $10^{-6}$, respectively. Fig. 2 illustrates the streamlines and isotherms for $Ra^*=100$ ($Ra^*=RaDa$) and $\phi=0.6$. From Fig. 2 we can observe that, for the same $Ra^*$, as $Da$ decreases, the velocity and thermal boundary layers near the hot (left) and cold (right) vertical walls become thinner. As $Da$ increases to $10^{-2}$, more convective mixing occurs inside the cavity, and the streamlines and isotherms are somewhat similar to those of the pure fluid cases. Those observed phenomena from the flow and temperature fields agree well with Refs. [5, 20]. To quantify the results, the average Nusselt numbers of the left vertical wall are calculated and listed in Table 2. The numerical results given by Nithiarasu *et al.* [5] using a finite element method and the numerical solutions obtained by Guo and Zhao [20] using a LBGK model are also listed in Table 2 for comparison. To sum up, our results agree well with the well-documented numerical results in previous studies.

4.2 Thermal convection in a porous cavity with isothermally cooled walls and internal heat generation

In this subsection, to test the applicability of the present MRT-LB model for convection flows in porous media with internal heat generation, we apply it to simulate thermal convection flow in a porous cavity with isothermally cooled walls in the presence of internal heat generation [47]. The geometry and boundary conditions are shown in Fig. 3. The four walls of the cavity ($A=1$) are maintained at temperature $T_c$. $Q'''$ is the internal heat source term, and the temperature difference $\Delta T$ is defined as $\Delta T = Q'''L^2/\alpha_e$. In this test, the reference temperature $T_0=T_c$, $Pr$ is set to be 7, $Ra=0$, and $Ra_I=6.4\times10^5$. The dimensionless relaxation time $\tau_\upsilon$ is determined by $\tau_\upsilon = 0.5 + MaJL\sqrt{Pr}/(c_s\delta_t\sqrt{Ra_I})$. A grid size of $120\times120$ is employed in simulations.

Numerical simulations using the present MRT-LB model are carried out based on the Brinkman-extended Darcy model, i.e., $\phi=1$, $F_\phi=0$. The streamlines and isotherms for different $Da$ are shown in Fig. 4. These plots are in good agreement with those reported in Ref. [47]. To quantify the comparisons, the maximum dimensionless stream function $\psi_{max}$ (normalized with $L\sqrt{g\beta\Delta TL}$) and the maximum dimensionless temperature $\theta_{max}$ ($\theta=(T-T_c)/\Delta T$) are calculated and included in Table 3 together with the data of Ref. [47]. From Table 3 we can observe that, the maximum dimensionless stream function $\psi_{max}$ decreases as $Da$ decreases, but the maximum dimensionless temperature $\theta_{max}$ increases. As shown, our results are in excellent agreement with the solutions reported in the literature.

4.3 Thermal convection in a porous cavity with internal heat generation

In this subsection, we apply the present MRT-LB model to study thermal convection flow in a porous cavity in the presence of internal heat generation, which has been investigated in Ref. [6]. The computational domain and boundary conditions are the same as natural convection problem in porous media (see Fig. 1), but a volumetric internal heat source $Q'''$ is applied in the domain. In this test, $Pr$ is fixed at $0.7$, and the grid sizes of $150\times150$ for $Ra=10^5$, and $200\times200$ for $Ra=10^6$ are employed.

The streamlines and isotherms for different $Ra_I$ with $Ra=10^5$, $\phi=0.4$, and $Da=10^{-2}$ are shown in Fig. 5. At $Ra_I=10^3$, an elliptic vortex appears in the cavity, and the heat transfer is dominated by convection due to the external sidewall-heating. As $Ra_I$ increases to $10^7$ (see Fig. 5c), two counter-rotating vortices appears in the cavity, and in the left part of the cavity, the isotherms curve in the opposite direction as compared with those in Figs. 5a and 5b.

Fig. 6 illustrates the streamlines and isotherms for different $Ra_I$ with $Ra=10^5$, $\phi=0.4$, and

$Da = 10^{-4}$. As $Da$ decreases to $10^{-4}$, the strength of the flow field reduces due to the low permeability of the porous media (compared with Fig. 5 for $Da = 10^{-2}$). At $Ra_I = 10^3$, the heat transfer in the cavity is still dominated by the external sidewall-heating, and a weak-convection structure can be observed from the isotherms. At $Ra_I = 10^7$, the flow field is governed by the internal-heating, and the heat transfer is dominated by convection due to internal-heating effect.

Fig. 7 illustrates the streamlines and isotherms for different $Da$ with $Ra = 10^6$, $Ra_I = 10^7$, and $\phi = 0.4$. For $Ra_I/Ra \geq 10$, the intensity of the internal-heating is much stronger than the external sidewall-heating, and the heat transfer from hot wall to the interior area has been suppressed by heat transfer from interior area to the hot wall. Furthermore, in the internal-heating dominated cases, as $Da$ decreases, the maximum dimensionless temperature $\theta_{max}$ increases. These observations from Figs. 5-7 agree well with those reported in Ref. [6].

To quantify the results, the maximum dimensionless temperature $\theta_{max}$ and the average Nusselt numbers of the hot wall are measured and included in Table 4. The numerical results of Ref. [6] using a finite element method are also included in Table 4 for comparison. From Table 4 we can see that, the present results are in quantitative agreement with those reported in Ref. [6].

## 5. Conclusions

In this paper, a MRT-LB model has been proposed for simulating incompressible thermal flows in porous media at the REV scale. The proposed MRT-LB model is constructed based on the MRT and DDF frameworks. The MRT-LB model consists of two MRT-LB equations: an incompressible D2Q9 MRT-LB equation is employed to solve the flow field, while a D2Q5 MRT-LB equation with a source term is used to solve the temperature field. The key point of the present MRT-LB model is that the effect of the porous media is incorporated into the MRT-LB model via introducing the porosity into the equilibrium moments and adding a forcing term, which accounts for the linear and nonlinear drag forces of the solid matrix, to the MRT-LB equation of the flow field in the moment space.

Numerical simulations of several 2D convection problems, including natural convection in a porous cavity, thermal convection in a porous cavity with isothermally cooled walls and internal heat generation, and thermal convection in a porous cavity with internal heat generation, have been performed to validate the proposed MRT-LB model. The results predicted by the present MRT-LB model agree well quantitatively with those reported in the literature.

**Acknowledgements**

This work was supported by the National Key Basic Research Program of China (973 Program) (2013CB228304), the National Natural Science Foundation of China (No. 51176155) and the Research Project of Chinese Ministry of Education (No.113055A).

**Appendix: Chapman-Enskog analysis**

The Chapman-Enskog analysis is employed to derive the macroscopic equation from the present D2Q5 MRT-LB equation. To this end, the following multiscale expansions in time and space are introduced [8, 39]:

$$g_i = g_i^{(0)} + \epsilon g_i^{(1)} + \epsilon^2 g_i^{(2)} + \cdots, \tag{A1a}$$

$$\frac{\partial}{\partial t} = \epsilon \frac{\partial}{\partial t_1} + \epsilon^2 \frac{\partial}{\partial t_2}, \tag{A1b}$$

$$\nabla = \epsilon \nabla_1, \quad \Psi = \epsilon \Psi_1, \quad Q''' = \epsilon Q_1''', \tag{A1c}$$

where $\epsilon$ is an expansion parameter.

Expanding $\mathbf{g}(\mathbf{x}_k + \mathbf{e}\delta_t, t_n + \delta_t)$ in Eq. (25) as a Taylor series about $\mathbf{x}_k$ and $t_n$, and substituting Eq. (A1) into the resulting equation, then multiplying by the transformation matrix $\mathbf{N}$, we can derive the following equations (in the moment space) in $\epsilon$ as

$$\epsilon^0: \quad \mathbf{n}^{(0)} = \mathbf{n}^{(eq)}, \tag{A2a}$$

$$\epsilon^1: \quad \left(\frac{\partial}{\partial t_1} + \mathbf{E}_\alpha \partial_{\alpha 1}\right)\mathbf{n}^{(0)} = -\mathbf{\Theta}\mathbf{n}^{(1)} + \mathbf{\Psi}_1, \tag{A2b}$$

$$\epsilon^2: \quad \frac{\partial \mathbf{n}^{(0)}}{\partial t_2} + \left(\frac{\partial}{\partial t_1} + \mathbf{E}_\alpha \partial_{\alpha 1}\right)\left(\mathbf{I} - \frac{1}{2}\mathbf{\Theta}\right)\mathbf{n}^{(1)} = -\mathbf{\Theta}\mathbf{n}^{(2)} - \frac{\delta_t}{2}\left(\frac{\partial}{\partial t_1} + \mathbf{E}_\alpha \partial_{\alpha 1}\right)\mathbf{\Psi}_1, \tag{A2c}$$

where $\mathbf{E}_\alpha = \mathbf{N}(\mathbf{e}_{i\alpha}\tilde{\mathbf{I}})\mathbf{N}^{-1}$ and $\mathbf{n}^{(1)} = \left(0, n_1^{(1)}, n_2^{(1)}, n_3^{(1)}, n_4^{(1)}\right)^T$.

Writing out the equations of Eq. (A2b), we obtain

$$\frac{\partial T}{\partial t_1} + \frac{\partial}{\partial x_1}\left(\frac{u_x T}{\sigma}\right) + \frac{\partial}{\partial y_1}\left(\frac{u_y T}{\sigma}\right) = \frac{Q_1'''}{\sigma}, \tag{A3a}$$

$$\frac{\partial}{\partial t_1}\left(\frac{u_x T}{\sigma}\right) + \frac{\partial}{\partial x_1}\left(\frac{2}{5}T + \frac{\varpi}{10}T\right) = -\zeta_1 n_1^{(1)}, \tag{A3b}$$

$$\frac{\partial}{\partial t_1}\left(\frac{u_y T}{\sigma}\right) + \frac{\partial}{\partial y_1}\left(\frac{2}{5}T + \frac{\varpi}{10}T\right) = -\zeta_2 n_2^{(1)}, \tag{A3c}$$

$$\frac{\partial(\varpi T)}{\partial t_1} + \frac{\partial}{\partial x_1}\left(\frac{u_x T}{\sigma}\right) + \frac{\partial}{\partial y_1}\left(\frac{u_y T}{\sigma}\right) = -\zeta_3 n_3^{(1)} + \frac{\varpi Q_1'''}{\sigma}, \tag{A3d}$$

$$\frac{\partial}{\partial x_1}\left(\frac{u_x T}{\sigma}\right) + \frac{\partial}{\partial y_1}\left(-\frac{u_y T}{\sigma}\right) = -\zeta_4 n_4^{(1)}. \tag{A3e}$$

The zeroth-order conserved moment of Eq. (A2c) is

$$\frac{\partial T}{\partial t_2} + \frac{\partial}{\partial x_1}\left[\left(1 - \frac{\zeta_1}{2}\right)n_1^{(1)}\right] + \frac{\partial}{\partial y_1}\left[\left(1 - \frac{\zeta_2}{2}\right)n_2^{(1)}\right] = -\frac{\delta_t}{2}\frac{\partial}{\partial t_1}\left(\frac{Q_1'''}{\sigma}\right). \tag{A4}$$

According to Eqs. (A3b) and (A3c), we have

$$n_1^{(1)} = -\frac{1}{\zeta_1}\left[\frac{\partial}{\partial t_1}\left(\frac{u_x T}{\sigma}\right) + \frac{(4+\varpi)}{10}\frac{\partial T}{\partial x_1}\right], \tag{A5a}$$

$$n_2^{(1)} = -\frac{1}{\zeta_2}\left[\frac{\partial}{\partial t_1}\left(\frac{u_y T}{\sigma}\right) + \frac{(4+\varpi)}{10}\frac{\partial T}{\partial y_1}\right]. \tag{A5b}$$

For incompressible flows, $\partial_{t_1}(u_x T/\sigma)$ and $\partial_{t_1}(u_y T/\sigma)$ in Eq. (A5) can be neglected [20], then Eq. (A5) can be written as

$$n_1^{(1)} = -\frac{1}{\zeta_1}\frac{(4+\varpi)}{10}\frac{\partial T}{\partial x_1}, \tag{A6a}$$

$$n_2^{(1)} = -\frac{1}{\zeta_2}\frac{(4+\varpi)}{10}\frac{\partial T}{\partial y_1}. \tag{A6b}$$

Substituting Eq. (A6) into Eq. (A5), then combining the resulting equation with Eq. (A3a) at the $t_1$ and $t_2$ time scales, we get

$$\frac{\partial T}{\partial t}+\frac{\partial}{\partial x}\left(\frac{u_x T}{\sigma}\right)+\frac{\partial}{\partial y}\left(\frac{u_y T}{\sigma}\right)=\frac{\epsilon(4+\varpi)}{10}\left[\left(\frac{1}{\zeta_1}-\frac{1}{2}\right)\frac{\partial^2 T}{\partial x^2}+\left(\frac{1}{\zeta_2}-\frac{1}{2}\right)\frac{\partial^2 T}{\partial y^2}\right]+\frac{Q'''}{\sigma}-\epsilon\frac{\delta_t}{2}\frac{\partial}{\partial t_1}\left(\frac{Q'''}{\sigma}\right). \tag{A7}$$

Assuming that $\sigma$ does not change with space and time, in the incompressible limit $\nabla \cdot \mathbf{u}=0$, we obtain the following macroscopic equation

$$\sigma\frac{\partial T}{\partial t}+\mathbf{u}\cdot\nabla T = \nabla\cdot(\alpha_e \nabla T)+Q'''-\epsilon\frac{\delta_t}{2}\frac{\partial Q'''}{\partial t_1}, \tag{A8}$$

with $\epsilon=\delta_t$ and $\zeta_1=\zeta_2=1/\tau_T$, $\alpha_e$ is the effective thermal diffusivity and given by

$$\alpha_e = \frac{\sigma(4+\varpi)}{10}\left(\tau_T-\frac{1}{2}\right)\delta_t. \tag{A9}$$

**References**

[1] P. Cheng, Heat transfer in geothermal systems, Adv. Heat Transfer 14 (1978) 1-105.

[2] D.A. Nield, A. Bejan, Convection in Porous Media, third ed., Springer, New York, 2006.

[3] C. Beckermann, R. Viskanta, S. Ramadhyani, Natural convection in vertical enclosures containing simultaneously fluid and porous layers, J. Fluid Mech. 186 (1988) 257-284.

[4] G. Lauriat, V. Prasad, Non-Darcian effects on natural convection in a vertical porous enclosure, Int. J. Heat Mass Transfer 32 (1989) 2135-2148.

[5] P. Nithiarasu, K.N. Seetharamu, T. Sundararajan, Natural convective heat transfer in a fluid

saturated variable porosity medium, Int. J. Heat Mass Transfer 40 (1997) 3955-3967.

[6] T.C. Jue, Analysis of thermal convection in a fluid-saturated porous cavity with internal heat generation, Heat Mass Transfer 40 (2003) 83-89.

[7] U. Frisch, B. Hasslacher, Y. Pomeau, Lattice-gas automata for the Navier-Stokes equation, Phys. Rev. Lett. 56 (1986) 1505.

[8] S. Chen, G. D. Doolen, Lattice Boltzmann method for fluid flows, Annu. Rev. Fluid Mech. 30 (1998) 329-364.

[9] Y. Gan, A. Xu, G. Zhang, Y. Li, Lattice Boltzmann study on Kelvin-Helmholtz instability: Roles of velocity and density gradients, Phys. Rev. E 83 (2011) 056704.

[10] Q. Li, Y.L. He, Y. Wang, W.Q. Tao, Coupled double-distribution-function lattice Boltzmann method for the compressible Navier-Stokes equations, Phys. Rev. E 76 (2007) 056705.

[11] Y.L. He, Y. Wang, Q. Li, Lattice Boltzmann Method: Theory and Applications, Science Press, Beijing, 2009.

[12] S. Succi, The Lattice Boltzmann Equation for Fluid Dynamics and Beyond, Clarendon Press, Oxford, 2001; S. Succi, Lattice Boltzmann across scales: from turbulence to DNA translocation, Eur. Phys. J. B 64 (2008) 471-479.

[13] S. Succi, E. Foti, F. Higuera, Three-dimensional flows in complex geometries with the lattice Boltzmann method, Europhys. Lett. 10 (1989) 433.

[14] G.H. Tang, W.Q. Tao, Y. L. He, Gas slippage effect on microscale porous flow using the lattice Boltzmann method, Phys. Rev. E 72 (5) (2005) 056301.

[15] Q. Kang, P.C. Lichtner, D. Zhang, An improved lattice Boltzmann model for multicomponent reactive transport in porous media at the pore scale, Water Resour. Res. 43 (12) (2007).

[16] O. Dardis, J. McCloskey, Lattice Boltzmann scheme with real numbered solid density for the simulation of flow in porous media, Phys. Rev. E 57 (1998) 4834.

[17] N.S. Martys, Improved approximation of the Brinkman equation using a lattice Boltzmann method, Phys. Fluids 6 (2001) 1807.

[18] Q. Kang, D. Zhang, S. Chen, Unified lattice Boltzmann method for flow in multiscale porous media, Phys. Rev. E 66 (2002) 056307.

[19] Z. Guo, T.S. Zhao, Lattice Boltzmann model for incompressible flows through porous media, Phys. Rev. E 66 (2002) 036304.

[20] Z. Guo, T.S. Zhao, A lattice Boltzmann model for convection heat transfer in porous media, Numer. Heat Transfer B 47 (2005) 157-177.

[21] T. Seta, E. Takegoshi, K. Okui, Lattice Boltzmann simulation of natural convection in porous media, Math. Comput. Simulat. 72 (2006) 195-200.

[22] P.L. Bhatnagar, E.P. Gross, M. Krook, A model for collision processes in gases. I. Small amplitude processes in charged and neutral one-component systems, Phys. Rev. 94 (1954) 511-525.

[23] F. J. Higuera, S. Succi, R. Benzi, Lattice gas dynamics with enhanced collisions, Europhys. Lett. 9 (1989) 345.

[24] D. d'Humières, Generalized lattice-Boltzmann equations, in: B.D. Shizgal, D.P. Weaver (Eds.), Rarefied Gas Dynamics: Theory and Simulations, in: Prog. Astronaut. Aeronaut., Vol. 159, AIAA, Washington, DC, 1992, pp. 450-458.

[25] P. Lallemand, L.-S. Luo, Theory of the lattice Boltzmann method: Dispersion, dissipation, isotropy, Galilean invariance, and stability, Phys. Rev. E 61 (2000) 6546.

[26] D. d'Humières, I. Ginzburg, M. Krafczyk, P. Lallemand, L.-S. Luo, Multiple–relaxation–time lattice Boltzmann models in three dimensions, Philos. Trans. R. Soc. Lond. A 360 (2002) 437-451.

[27] P. Lallemand, L.-S. Luo, Theory of the lattice Boltzmann method: Acoustic and thermal properties in two and three dimensions, Phys. Rev. E 63 (2003) 036706.

[28] Q. Li, Y.L. He, G.H. Tang, W.Q. Tao, Improved axisymmetric lattice Boltzmann scheme, Phys. Rev. E 81 (2010) 056707.

[29] Q. Li, Y.L. He, G.H. Tang, W.Q. Tao, Lattice Boltzmann modeling of microchannel flows in the transition flow regime, Microfluid. Nanofluid. 10 (2011) 607-618.

[30] M.E. MaCracken, J. Abraham, Multiple-relaxation-time lattice-Boltzmann model for multiphase flow, Phys. Rev. E 71 (2005) 036701.

[31] Q. Li, K.H. Luo, X.J. Li, Lattice Boltzmann modeling of multiphase flows at large density ratio with an improved pseudopotential model, Phys. Rev. E, 87 (2013) 053301.

[32] A. Mezrhab, M.A. Moussaoui, M. Jami, H. Naji, M. Bouzidi, Double MRT thermal lattice Boltzmann method for simulating convective flows, Phys. Lett. A 374 (2010) 3499-3507.

[33] J. Wang, D. Wang, P. Lallemand, L.-S. Luo, Lattice Boltzmann simulations of thermal convective flows in two dimensions, Comput. Math. Appl. 66 (2013) 262-286.

[34] C.T. Hsu, P. Cheng, Thermal dispersion in a porous medium, Int. J. Heat Mass Transfer 33 (1990) 1587-1597.

[35] S. Ergun, Fluid flow through packed columns, Chem. Eng. Prog. 48 (1952) 89-94.

[36] K. Vafai, Convective flow and heat transfer in variable-porosity media, J. Fluid Mech. 147 (1984) 233-259.

[37] Y.H. Qian, D. d'Humeier, P. Lallemand, Lattice BGK models for Navier-Stokes equation,

Europhys. Lett. 17 (1992) 479-484.

[38] Z. Guo, B. Shi, N. Wang, Lattice BGK model for incompressible Navier-Stokes equation, J. Comput. Phys. 165 (2000) 288-306.

[39] S. Chapman and T. G. Cowling, The Mathematical Theory of Non-Uniform Gases, Cambridge University Press, London, 1970.

[40] R. Du, B. Shi, X. Chen, Multi-relaxation-time lattice Boltzmann model for incompressible flow, Phys. Lett. A 359 (2006) 564-572.

[41] T. Seta, Implicit temperature-correction-based immersed-boundary thermal lattice Boltzmann method for the simulation of natural convection, Phys. Rev. E 87, (2013) 063304.

[42] I. Rasin, S. Succi, W. Miller, A multi-relaxation lattice kinetic method for passive scalar diffusion, J. Comput. Phys. 206 (2005) 453-462.

[43] Z.L. Guo, C.G. Zheng, B.C. Shi, Non-equilibrium extrapolation method for velocity and pressure boundary conditions in the lattice Boltzmann method, Chin. Phys. 11 (2002) 366.

[44] H.N. Dixit, V. Babu, Simulation of high Rayleigh number natural convection in a square cavity using the lattice Boltzmann method, Int. J. Heat Mass Transfer 49 (2006) 727-739.

[45] Q. Li, Y.L. He, Y. Wang, G.H. Tang, An improved thermal lattice Boltzmann model for flows without viscous heat dissipation and compression work, Int. J. Mod. Phys. C 19 (2008) 125-150.

[46] M. Hortmann, M. Perić, G. Scheuerer, Finite volume multigrid prediction of laminar natural convection: Bench-mark solutions, Int. J. Numer. Meth. Fluids 11 (1990) 189-207.

[47] K.M. Khanafer, A.J. Chamkha, Hydromagnetic natural convection from an inclined porous square enclosure with heat generation, Numer. Heat Transfer A 33 (1998) 891-910.

**Figure Captions**

Fig. 1. Geometry and boundary conditions of example 1.

Fig. 2. Streamlines (left), and isotherms (right) for $Ra_I = 0$, $\phi = 0.4$, and $Pr = 1$.

Fig. 3. Geometry and boundary conditions of example 2.

Fig. 4. Streamlines (left), and isotherms (right) for $Ra = 0$, $Ra_I = 6.4 \times 10^5$, $\phi = 1$, $F_\phi = 0$, and $Pr = 7$.

Fig. 5. Streamlines (left), and isotherms (right) for $Ra = 10^5$, $\phi = 0.4$, and $Da = 10^{-2}$.

Fig. 6. Streamlines (left), and isotherms (right) for $Ra = 10^5$, $\phi = 0.4$, and $Da = 10^{-4}$.

Fig. 7. Streamlines (left), and isotherms (right) for $Ra = 10^6$, $Ra_I = 10^7$, and $\phi = 0.4$.

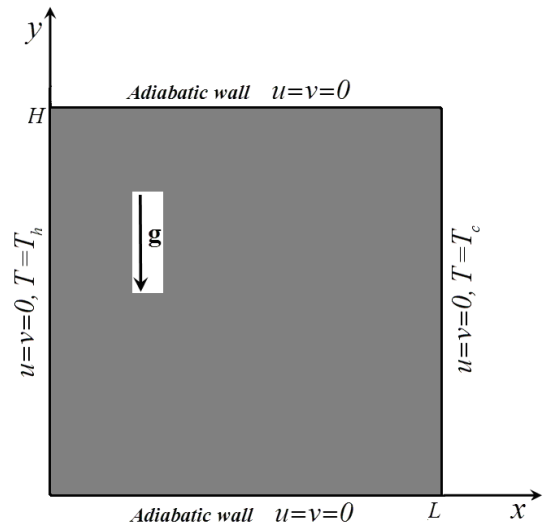

Fig. 1. Geometry and boundary conditions of example 1.

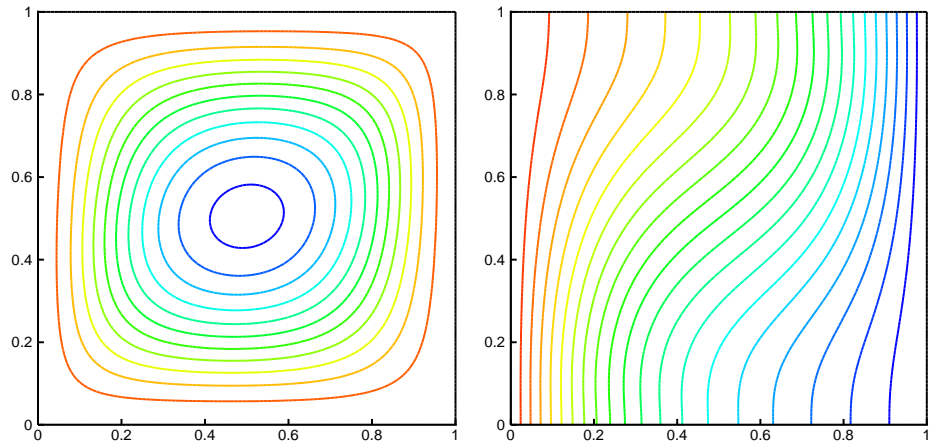

(a) $Da = 10^{-2}$, $Ra = 10^4$

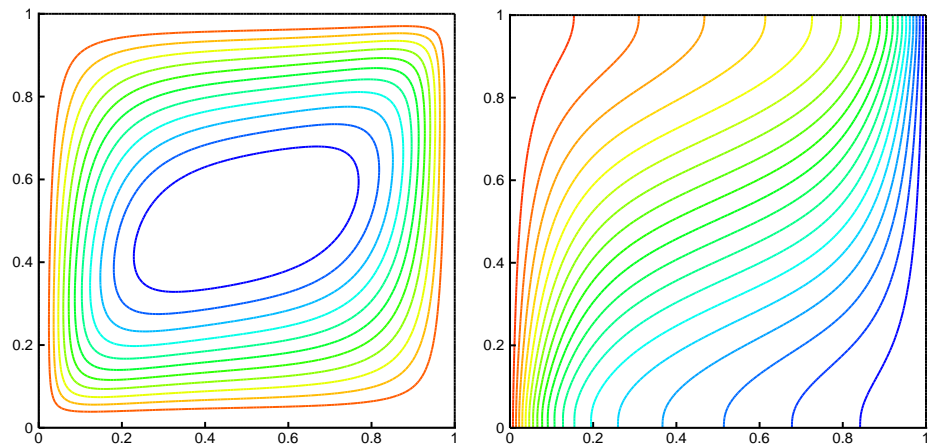

(a) $Da = 10^{-4}$, $Ra = 10^6$

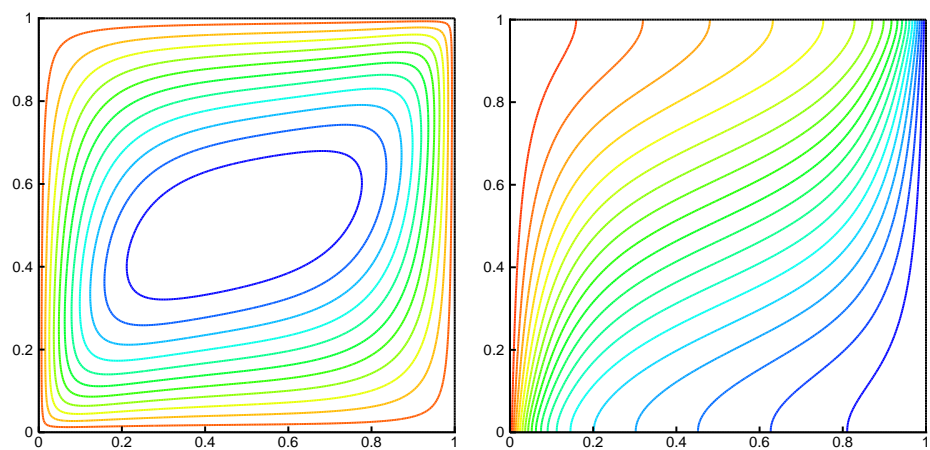

(a) $Da = 10^{-6}$, $Ra = 10^8$

Fig. 2. Streamlines (left), and isotherms (right) for $\phi = 0.4$ and $Pr = 1$.

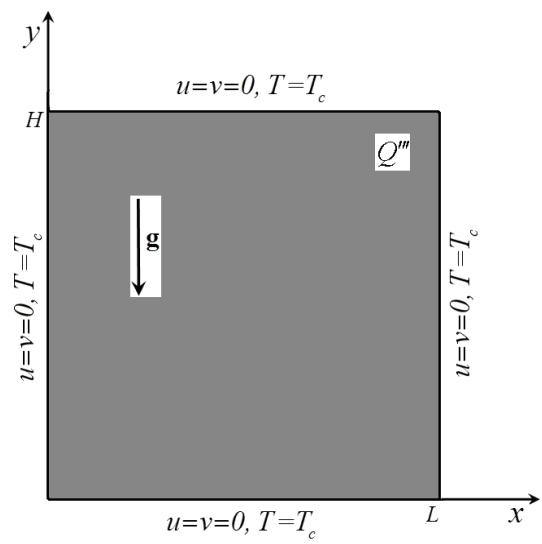

Fig. 3. Geometry and boundary conditions of example 2.

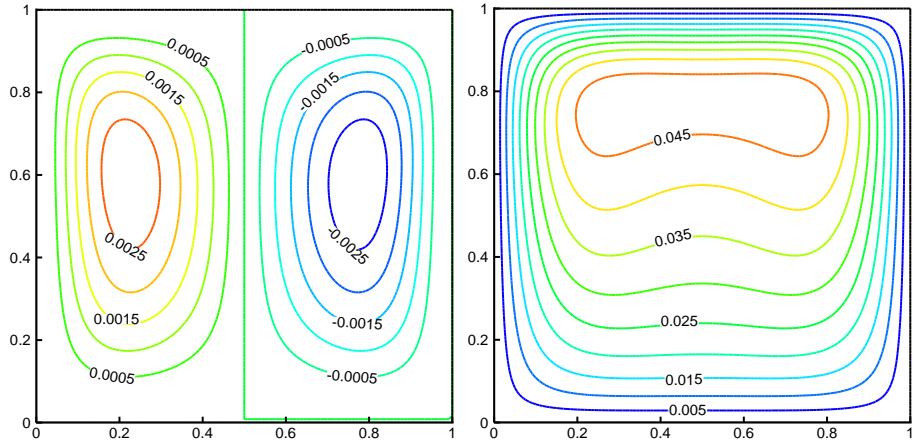

(a) $Da = \infty$

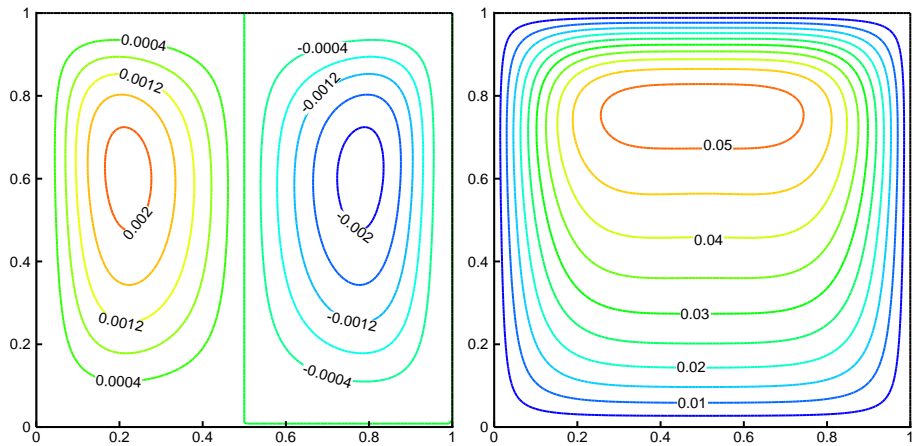

(b) $Da = 10^{-2}$

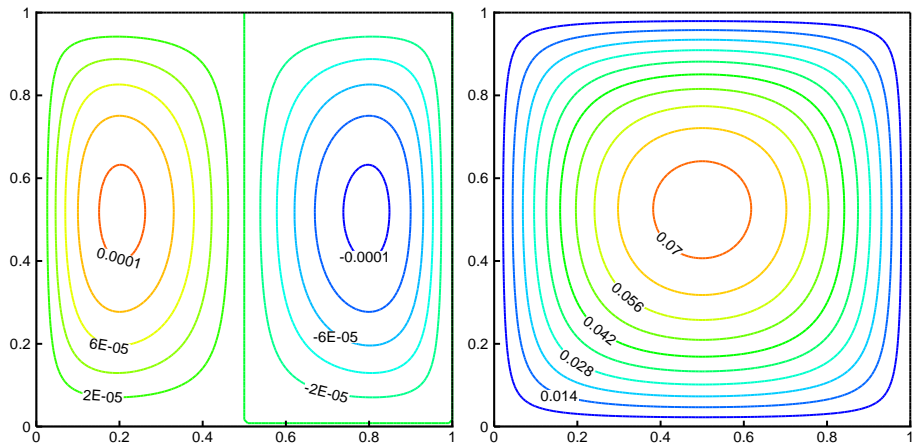

(c) $Da = 10^{-4}$

Fig. 4. Streamlines (left), and isotherms (right) for $Ra = 0$, $Ra_I = 6.4 \times 10^5$, $\phi = 1$, $F_\phi = 0$, and $Pr = 7$.

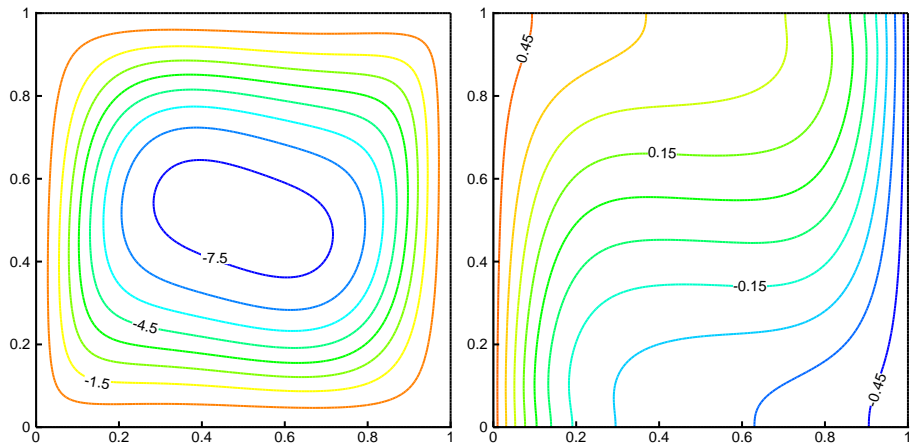

(a) $Ra_I = 10^3$

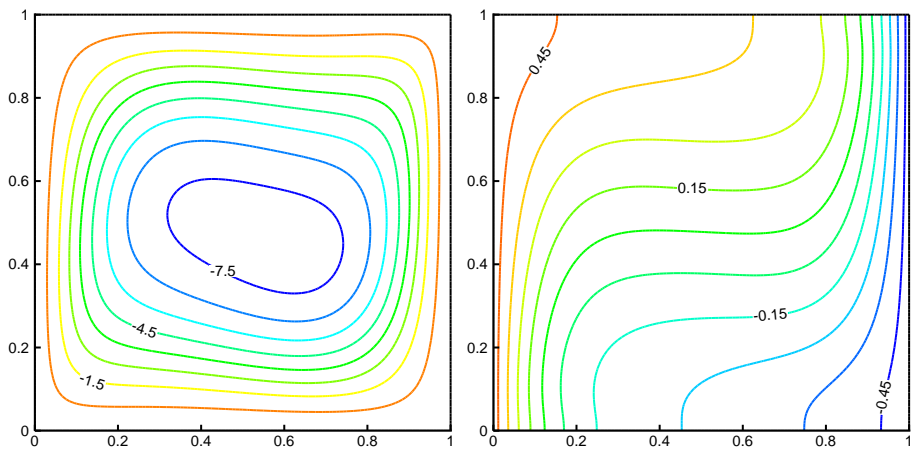

(b) $Ra_I = 10^5$

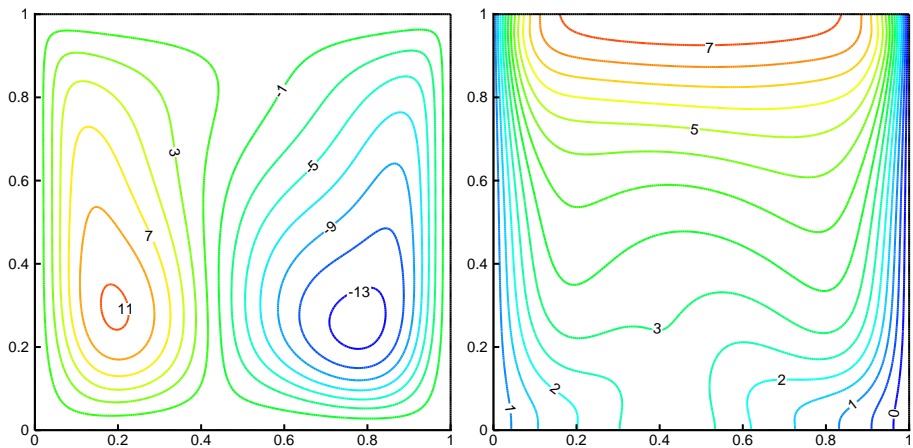

(c) $Ra_I = 10^7$

Fig. 5. Streamlines (left), and isotherms (right) for $Ra = 10^5$, $\phi = 0.4$, and $Da = 10^{-2}$.

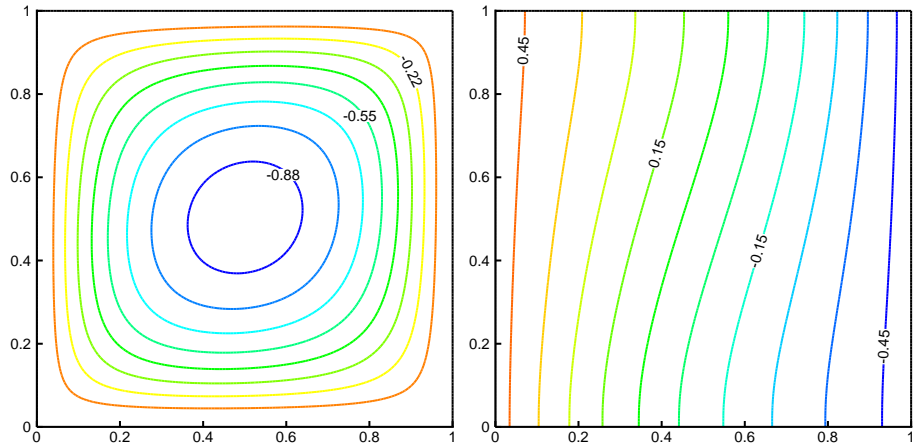

(a) $Ra_I = 10^3$

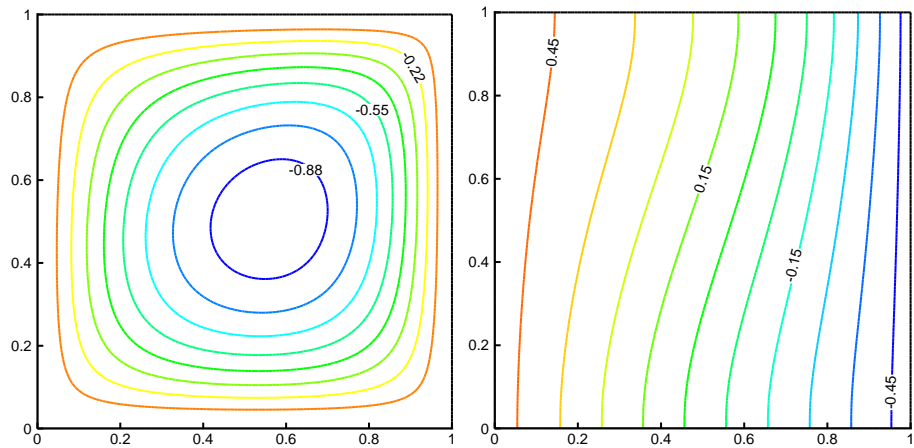

(b) $Ra_I = 10^5$

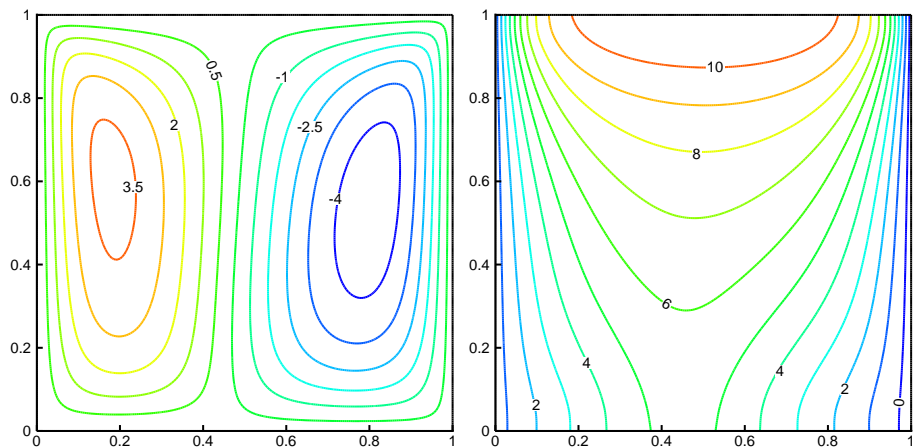

(c) $Ra_I = 10^7$

Fig. 6. Streamlines (left), and isotherms (right) for $Ra = 10^5$, $\phi = 0.4$, and $Da = 10^{-4}$.

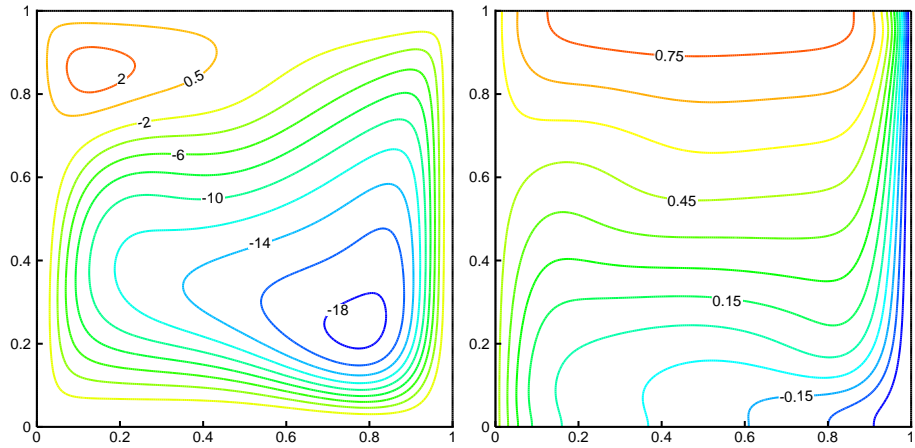

(a) $Da = 10^{-2}$

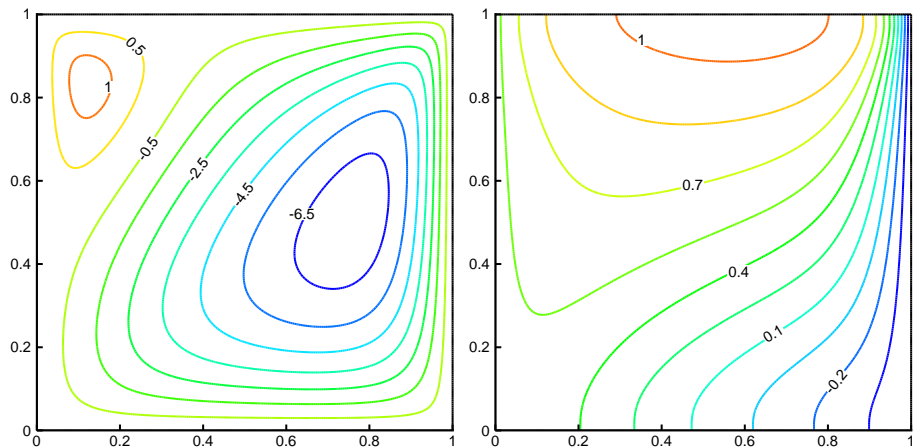

(b) $Da = 10^{-4}$

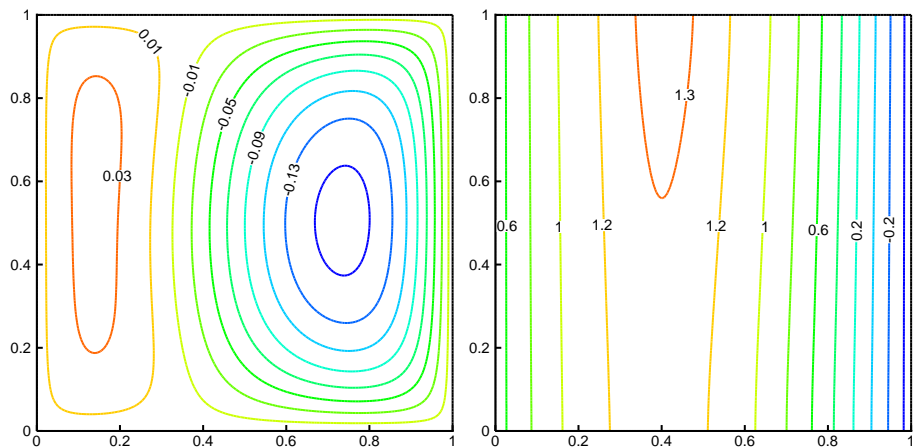

(c) $Da = 10^{-6}$

Fig. 7. Streamlines (left), and isotherms (right) for $Ra = 10^6$, $Ra_I = 10^7$, and $\phi = 0.4$.

**Table Captions**

Table 1 Comparisons of the average Nusselt numbers ($\phi = 0.9999$, $Ra_I = 0$, $Da = 10^8$, $Pr = 0.71$).

Table 2 Comparisons of the average Nusselt numbers for moderate $\phi$ and $Da$ ($Pr = 1.0$, $Ra_I = 0$).

Table 3 Comparisons of the present results for $\psi_{max}$ and $\theta_{max}$ with those reported in Ref. [47] ($Ra = 0$, $Ra_I = 6.4 \times 10^5$, $\phi = 1$, $F_\phi = 0$, $Pr = 7$).

Table 4 Comparisons of the present results for $\overline{Nu}$ and $\theta_{max}$ with those reported in Ref. [6] ($Pr = 0.7$).

Table 1 Comparisons of the average Nusselt numbers ($\phi = 0.9999$, $Ra_I = 0$, $Da = 10^8$, $Pr = 0.71$).

| $Ra$ | Ref. [44] | Ref. [45] | Ref. [46] | Present |
|---|---|---|---|---|
| $10^3$ | 1.121 | 1.1169 | - | 1.1160 |
| $10^4$ | 2.286 | 2.2452 | 2.2448 | 2.2447 |
| $10^5$ | 4.546 | 4.5219 | 4.5216 | 4.5197 |
| $10^6$ | 8.652 | 8.7926 | 8.8251 | 8.7881 |
| $10^7$ | 16.79 | - | - | 16.4386 |

Table 2 Comparisons of the average Nusselt numbers for moderate $\phi$ and $Da$

( $Pr = 1.0$, $Ra_I = 0$ ).

| $Da$ | $Ra$ | $\phi = 0.4$ | | | $\phi = 0.6$ | | |
|---|---|---|---|---|---|---|---|
| | | Ref. [5] | Ref. [20] | Present | Ref. [5] | Ref. [20] | Present |
| $10^{-2}$ | $10^3$ | 1.010 | 1.008 | 1.007 | 1.015 | 1.012 | 1.012 |
| | $10^4$ | 1.408 | 1.367 | 1.362 | 1.530 | 1.499 | 1.494 |
| | $10^5$ | 2.983 | 2.998 | 3.009 | 3.555 | 3.422 | 3.460 |
| $10^{-4}$ | $10^5$ | 1.067 | 1.066 | 1.067 | 1.071 | 1.068 | 1.069 |
| | $10^6$ | 2.550 | 2.603 | 2.630 | 2.725 | 2.703 | 2.733 |
| | $10^7$ | 7.810 | 7.788 | 7.808 | 8.183 | 8.419 | 8.457 |
| $10^{-6}$ | $10^7$ | 1.079 | 1.077 | 1.085 | 1.079 | 1.077 | 1.089 |
| | $10^8$ | 2.970 | 2.955 | 2.949 | 2.997 | 2.962 | 2.957 |
| | $10^9$ | 11.460 | 11.395 | 11.610 | 11.790 | 11.594 | 12.092 |

Table 3 Comparisons of the present results for $\psi_{max}$ and $\theta_{max}$ with those reported in Ref. [47] ($Ra = 0$, $Ra_I = 6.4 \times 10^5$, $\phi = 1$, $F_\phi = 0$, $Pr = 7$).

| $Da$ | | Ref. [47] | Present |
|---|---|---|---|
| $\infty$ | $\psi_{max}$ | $2.91 \times 10^{-3}$ | $2.86 \times 10^{-3}$ |
| | $\theta_{max}$ | $4.75 \times 10^{-2}$ | $4.79 \times 10^{-2}$ |
| $10^{-2}$ | $\psi_{max}$ | $2.21 \times 10^{-3}$ | $2.17 \times 10^{-3}$ |
| | $\theta_{max}$ | $5.22 \times 10^{-2}$ | $5.26 \times 10^{-2}$ |
| $10^{-4}$ | $\psi_{max}$ | $1.10 \times 10^{-4}$ | $1.06 \times 10^{-4}$ |
| | $\theta_{max}$ | $7.35 \times 10^{-2}$ | $7.34 \times 10^{-2}$ |

Table 4 Comparisons of the present results for $\overline{Nu}$ and $\theta_{max}$ with those reported in Ref. [6] ($Pr = 0.7$, $\phi = 0.4$).

| $Ra$ | $Ra_I$ | | $Da = \infty$, $\phi = 1$ | | $Da = 10^{-2}$ | | $Da = 10^{-4}$ | | $Da = 10^{-6}$ | |
|---|---|---|---|---|---|---|---|---|---|---|
| | | | $\overline{Nu}$ | $\theta_{max}$ | $\overline{Nu}$ | $\theta_{max}$ | $\overline{Nu}$ | $\theta_{max}$ | $\overline{Nu}$ | $\theta_{max}$ |
| $10^5$ | $10^3$ | Ref. [6] | 4.505 | 0.5 | 2.884 | 0.5 | 1.058 | 0.5 | 0.995 | 0.5 |
| | | Present | 4.538 | 0.5 | 2.903 | 0.5 | 1.061 | 0.5 | 0.995 | 0.5 |
| | $10^5$ | Ref. [6] | 4.021 | 0.5 | 2.401 | 0.5 | 0.571 | 0.5 | 0.506 | 0.5 |
| | | Present | 4.057 | 0.5 | 2.421 | 0.5 | 0.575 | 0.5 | 0.506 | 0.5 |
| | $10^7$ | Ref. [6] | -42.45 | 5.54 | -44.08 | 7.30 | -46.37 | 10.86 | -48.35 | 12.63 |
| | | Present | -41.35 | 5.52 | -43.41 | 7.23 | -45.08 | 10.76 | -48.31 | 13.27 |
| $10^6$ | $10^7$ | Ref. [6] | 4.129 | 0.66 | 1.242 | 0.81 | -1.789 | 1.01 | -3.93 | 1.32 |
| | | Present | 4.254 | 0.65 | 1.296 | 0.80 | -1.728 | 1.05 | -3.926 | 1.32 |